\documentclass[a4paper,fleqn]{cas-sc}

\usepackage[numbers]{natbib}

\def\tsc#1{\csdef{#1}{\textsc{\lowercase{#1}}\xspace}}
\tsc{WGM}
\tsc{QE}

\begin{document}
\let\WriteBookmarks\relax
\def\floatpagepagefraction{1}
\def\textpagefraction{.001}

\shorttitle{}    

\shortauthors{}  

\title [mode = title]{A statistical complexity measure can differentiate Go/NoGo trials during a visual-motor task using human electroencephalogram data}  

\tnotemark[1] 

\tnotetext[1]{} 
\author[1]{Francisco Leandro P. Carlos}
\author[1,2]{Maria Carla Navas}
\author[1]{Ícaro Rodolfo Soares Coelho Da Paz}
\author[1,3]{Helena Bordini de Lucas}
\author[4,5]{Maciel-Monteiro Ubirakitan}
\author[4]{Marcelo Cairr\~ao Ara\'ujo Rodrigues}
\author[5,6]{Moises Aguilar Domingo}
\author[7]{Eva Herrera-Guti\'errez}
\author[1]{Osvaldo A. Rosso}
\author[2]{Luz Bavassi}
\author[1]{Fernanda Selingardi Matias}

\cormark[1]
\cortext[1]{Corresponding author: fernanda@fis.ufal.br} 


\affiliation[1]{organization={Instituto de Física, Universidade Federal de Alagoas}, 
        city={Macei\'{o}, Alagoas 57072-970 }, 
        country={Brazil}}

\affiliation[2]{organization={Laboratorio de Neurociencias de la Memoria, IFIBYNE, Universidade de Buenos Aires},
    city={Buenos Aires}, 
    country={Argentina}}

\affiliation[3]{organization={Instituto de F\'{\i}sica Interdisciplinar y Sistemas Complejos, Universitat de les Illes Balears}, 
    city={Palma de Mallorca}, 
    country={Spain}}

\affiliation[4]{organization={Departamento de Fisiologia e Farmacologia, Universidade Federal de Pernambuco}, 
    city={Recife PE 50670-901}, 
    country={Brazil}}

\affiliation[5]{organization={Spanish Foundation for Neurometrics Development, Department of Psychophysics \& Psychophysiology}, 
    city={Murcia}, 
    country={Spain}}

\affiliation[6]{organization={Department of Human Anatomy and Psychobiology, Faculty of Psychology, University of Murcia},
    city={Murcia}, 
    country={Spain}}

\affiliation[7]{organ
ization={Department of Developmental and Educational Psychology, Faculty of Psychology, University of Murcia},
    city={Murcia}, 
    country={Spain}}

\begin{abstract}
Complexity is a ubiquitous concept in contemporary science and everyday life. A complex dynamical system is usually characterized by a blend of order and disorder, as well as emergent phenomena
that often span multiple temporal and spatial scales. The information processes related to different cognitive processes in the brain can be studied in light of statistical differences based on complexity measures of the electrophysiological time series from different trial types. Recently, it has been demonstrated that a symbolic information approach can be a valuable tool for discriminating response-related differences between Go and NoGo trials using the local field potential of brain regions in monkeys. The method shows significant differences between trial types earlier than the simple average of the electrical signals.
Here, we analyze human electroencephalogram data during a Go/NoGo task using information-theoretical quantifiers, including entropy and complexity. We employ the Bandt-Pompe symbolization methodology to determine a probability distribution function for each time series and then to calculate information theory indices. We show that a few channels,
especially a central and an occipital electrode, consistently differentiate Go/NoGo trials at the individual level. We also show that different trial types occupy separate regions in the complexity-entropy plane. Our method also captures specific time windows to better differentiate the trial type. Moreover, we show that our results are robust at the group level, considering the average activity of all subjects.
\end{abstract}


\begin{highlights}
\item 
\item 
\item 
\end{highlights}


\maketitle

\section{Introduction}
\label{sec:introduction}

Efficient performance during visuomotor cognitive tasks relies on the integrated processing of information by the visual and motor systems within the cerebral cortex \cite{Ledberg07}. Understanding how this information process occurs is one of the most significant questions in neuroscience. A few specific points can be addressed in this direction, such as identifying the statistical differences between two different tasks, determining when these differences are more pronounced, determining which timescales are more relevant for each process, and identifying which channels and brain regions are more engaged in these tasks.

Here, we employ a symbolic information approach~\cite{Bandt02} and information theory quantifiers
\cite{Shannon49,lopez1995statistical,Lamberti04, Rosso07, Zunino12, xiong2020complexity} to address these questions, analyzing EEG data from human subjects during a Go/NoGo task. The Go/NoGo task is a type of visual continuous performance neuropsychological test designed to study complex attentional functions, including response, inhibition, and sustained attention~\cite{harmony2009time}.

Over the last decades, integrated activity in the visual and motor systems during visual discrimination tasks has been widely investigated through the analysis of monkey Local Field Potential (LFP) data \cite{Bressler93, Liang00, Brovelli04, Salazar12, Dotson14}. Typically, the average evoked response potential (ERP) from monkey LFPs has been used to assess response-related differences between trial types, for example, in Go / NoGo tasks~\cite{Ledberg07}. The findings indicate that response-specific neural processing begins approximately $150$~ms after stimulus onset across broad cortical regions. The temporal evolution of task-related neural activity was analyzed to identify cortical areas and time intervals that showed significant differences between the two trial types.

Recently, it has been shown that the Bandt-Pompe symbolization methodology~\cite{Bandt02} applied to concatenated trials could provide more information than comparing the activation response by averaging brain activity~\cite{deLucas21}. A similar methodology has also been applied to intracranial human data to characterize cortical activity and to differentiate trial types during a visual search task~\cite{da2024symbolic}. 
The methodology employed here employs time causal quantifiers based on Information Theory (Shannon entropy, MPR-statistical complexity, and entropy-complexity plane) 
\cite{Shannon49,Lamberti04, Rosso07, Zunino12, xiong2020complexity}.
These quantifiers are evaluated using the Bandt-Pompe symbolization methodology~\cite{Bandt02}, which naturally incorporates the time-causal ordering provided by the time-series data into the corresponding associated probability distribution function (PDF). Different brain signals have been analyzed in the light of information theory indices in order to estimate time differences during phase synchronization~\cite{Montani15}, to show that complexity is maximized close to criticality in cortical states~\cite{lotfi2021statistical},
to distinguish cortical states using EEG data~\cite{rosso2006eeg}, MEG data~\cite{echegoyen2020permutation}, local field potential~\cite{jungmann2024state} as well as neuronal activity~\cite{Montani2015neuronas,Montani2014,deLuise2021network}.

In Sec.~\ref{methods}, we start by describing the experimental paradigm: the  EEG processing and acquisition, as well as the Go/NoGo task and the concatenated time series.
Then, in subsection~\ref{Symbolic representation of a time series: Bandt-Pompe approach}, we describe the symbolization technique called the Bandt-Pompe methodology to calculate the probability distribution function (PDF) associated with a time series. In subsection~\ref{Information theory quantifiers}, we introduce the information theory quantifiers: the Shannon entropy and the Jensen-Shannon statistical complexity based on the disequilibrium between the time series of interest and one with an equiprobable PDF. In Sec.~\ref{results}, we report our results, showing that the MPR complexity-entropy ($CxH$) plane can help visualize response-related differences between trial types. We then demonstrate that the Euclidean distance between two conditions in the $CxH$ plane serves as a reliable index for identifying channels and time windows that are more effective in distinguishing Go and NoGo trials. Finally, we compare the results at both the individual and group levels. Concluding remarks and a brief discussion of the significance of our findings for neuroscience are presented in Sec.~\ref{conclusion}.

\section{\label{methods}Methods}

\subsection{Human electroencephalogram data}

The electroencephalographic data recordings were carried out at the Spanish Foundation for Neurometrics Development (Murcia, Spain) center.
The entire experimental protocol was approved by the Commission of Bioethics of the University of
Murcia (UMU, process C25212072413233).
Part of the data analyzed here has been previously analyzed by Carlos et al.~\cite{carlos2020anticipated} in the context of anticipated synchronization ideas, using Granger causality measures to infer connectivity during the first $300$~ms of Go/NoGo trials.

The available dataset consists of EEG data from 11 volunteers 
(10 women, one man) right-handed, who gave informed consent to participate in the experiment.
The participants' ages ranged from 32 to 55 years, with a mean of 45.7 years and a standard deviation of 7.8. Each individual underwent evaluations by both a psychiatrist and a psychologist.
The exclusion criteria included a history of perinatal complications, head trauma resulting in loss of consciousness or neurological impairment, past seizures, use of medication or other substances within 24 hours prior to data collection, presence of psychotic symptoms in the six months preceding the study, and any diagnosed systemic or neurological disorders.

EEG data were recorded using a Mitsar 201M amplifier (Mitsar Ltd), a 19-channel system with an auricular reference. The signals were digitized at a sampling rate of 250 Hz. Data acquisition was carried out using the WinEEG software (Version 2.92.56). Electrodes were applied following the international 10-20 system, with conductive gel (ECI ELECTRO-GEL) ensuring proper contact.

\begin{figure}[h]
\centering
\includegraphics[width=0.8\linewidth]{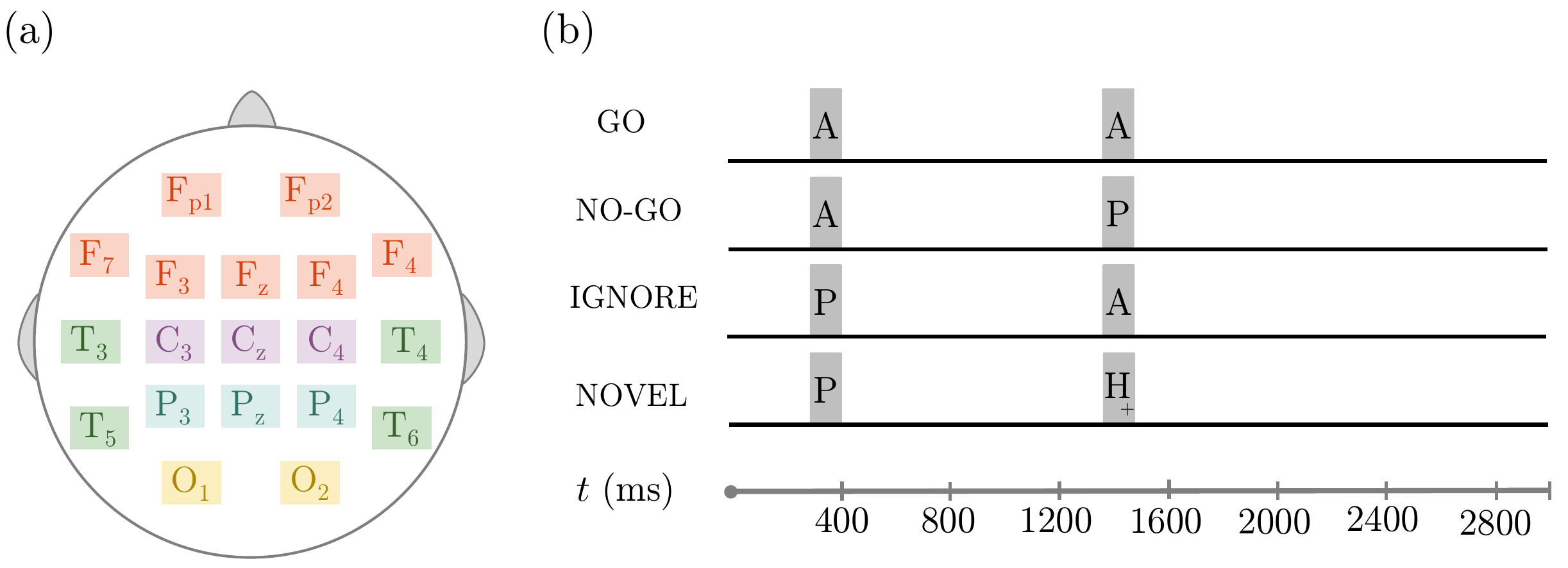}
\caption{
{\bf Experimental details.} 
(a) The EEG electrodes were positioned according to the international 10-20 system.
(b) Description of the Go/NoGo task based on three types of images: animals (A), plants (P), and people (H$_+$).
After a waiting window of 300~ms, two images were presented for 100 milliseconds each, with a 1000 ms inter-stimulus interval.
 Go trials consist of two similar stimuli: AA, and the participant should press a button as quickly as
possible. NoGo trials consist of a different pair of stimuli: AP and the subject should only wait for the subsequent trial (see Methods for more details).
}
\label{fig:task1}
\end{figure}

The montage includes three midline sites: F$_Z$, C$_Z$ and P$_Z$, and sixteen other sites:
$F_{P1}$, $F_{P2}$, $F_{7}$, $F_{8}$, $F_{3}$, $F_{4}$, $T_{3}$, $T_{4}$, $C_{3}$, $C_{4}$, $P_{3}$, $P_{4}$, $T_{5}$, $T_{6}$, $O_{1}$, and $O_{2}$. See an example of the headset in Fig.~\ref{fig:task1}(a). The electrode impedance was maintained $<5$~ K$\Omega$. EEG epochs with excessive amplitude ($>50$~$\mu$V) were automatically deleted. Finally, the EEG was analyzed by a specialist in neurophysiology to reject epochs with artifacts. 

\subsection{\label{subsec:task} Go/NoGo visual task}

The EEG data were recorded while subjects performed a visual continuous performance task (VCPT), also called a Go/NoGo task. Depending on the stimulus, the volunteer should press a button as quickly as possible.
Participants sat in an ergonomic chair 1.5 meters away
from a $17''$ plasma screen. Psytask software (Mitsar Systems) was used to present the images.
The VCPT consists of three types of stimuli: twenty images of animals (A), twenty images
of plants (P), and twenty images of people of different professions (H$_+$). 
Whenever H$_+$ was presented, a $20$~ms-long artificial sound tone frequency
was simultaneously produced. The tone frequencies range from 500 to $2500$~Hz, in intervals of $500$~Hz.
All stimuli were of equal size and brightness. 

In each trial, the first
stimuli were presented after a waiting window of $300$~ms, (see Fig.~\ref{fig:task1}(b)). 
The stimulus remains on the screen for $100$~ms, and after $1000$~ms a second stimulus appears on the screen and remains for $100$~ms. Therefore, we define $t_a=300$~ms, and $t_b=1400$~ms as the initial time of the first and second stimulus. 
Four different kinds of pairs of stimuli were employed: AA, AP, PP, and PH$_+$. 
The entire experiment consists of 400 trials (the four trial types were randomly distributed, and each one appeared 100 times).
The continuous set occurs when A is presented as the first stimulus, so the subject needs to prepare to respond. An AA pair corresponds to a Go task, and the participants are supposed to press a button as quickly as possible. An AP pair corresponds to a 
NoGo task, and the participants should
suppress the action of pressing the button. In this paper, we analyze only the continuous set and compare 100 Go trials and 100 NoGo trials from each volunteer.
The discontinuous set, in which P
is first presented, indicates that one should not respond (independently of the second stimulus). IGNORE task occurred with PP
pairs and NOVEL when PH$_+$ pairs appeared.
Participants were trained for about five minutes before beginning the experimental trials. They rested for
a few minutes when they reached the halfway point of the task. The experimental session lasted $\sim30$~min.

\begin{figure}
\centering
\includegraphics[width=0.8\linewidth]{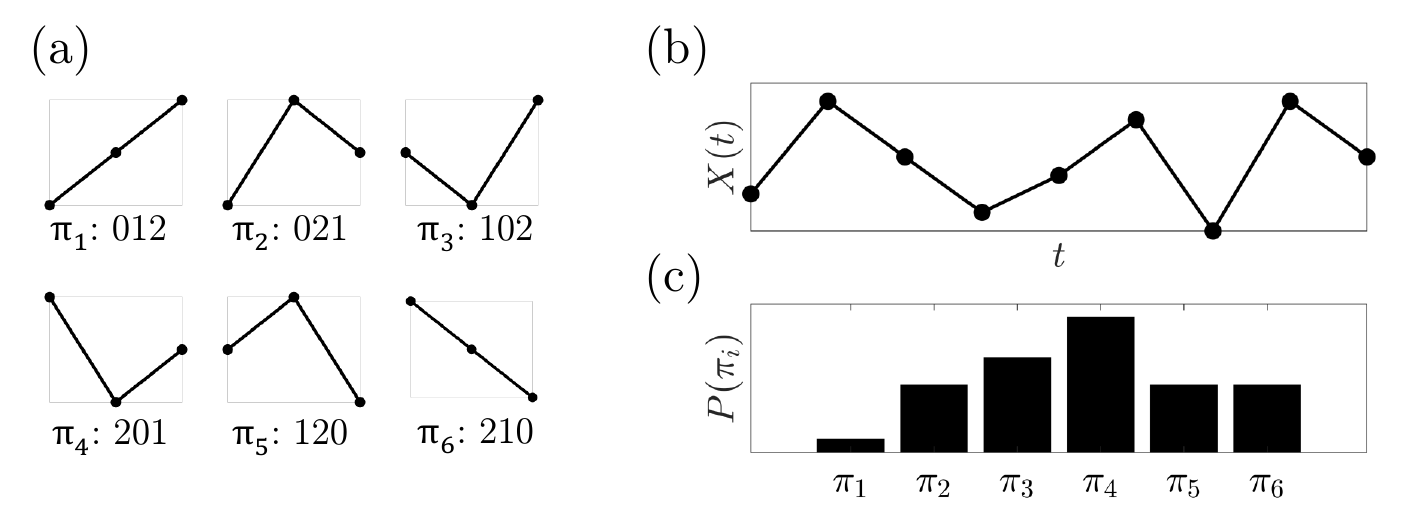}
\caption{\label{fig:DSAS} 
Characterizing the symbolic representation of time series.
(a) The six possible symbols associated with permutations $\pi_j$ for ordinal patterns of length $D=3$.
(b) Example of a straightforward time series $X(t)= {4, 9, 6, 3, 5, 8, 2, 9, 6}$ and (c) Illustrative probability distribution function (PDF) with the frequencies of each symbol.
}
\label{fig:symbol}
\end{figure}

\subsection{\label{subsec:timeseries}Time series}
Each trial takes $3000$~ms, measured every $4$~ms (sampling rate of $250$~Hz). 
We separate the trial into windows of 100~ms, starting every 20~ms. 
We analyze the data by concatenating all j-th windows of the same type of trials. This means that we concatenate 25 points from every trial and analyze the time series of 2500 (25 timepoints × 100 trials).
The $j=16$ window starts at $t_{16}^i=300$~ms and ends at $t_{16}^f=400$~ms. Therefore, it is exactly in which the first image is on the screen ($t_a=t_{16}^i=300$~ms), whereas the $j=71$ is the window in which the second image is on the screen (it starts at $t_b=t_{71}^i=1400$~ms). Unless otherwise stated, we will refer throughout the text only to the initial time of the j-th window as $t$ instead of $t_j^i$.

\subsection{\label{Symbolic representation of a time series: Bandt-Pompe approach}{Symbolic representation of a time series: Bandt-Pompe approach}}

To calculate any information-theory quantifier, one should obtain a probability distribution function (PDF) from a time series $X(t)$ representing the evolution dynamics of the system under study. Let $X(t)\equiv \{x_t;t=1,2, \dots ,M\}$, be the time series representing a set of $M$ measures of the observable $X$. 
Here, we use a symbolic representation of a time series introduced by Bandt-Pompe~\cite{Bandt02} for evaluating the PDF. This symbolization technique consists of extracting the ordinal patterns of
length $D$, associated with each time $t$ of our time series, generated by $\textbf{s}(t)=(x_{t},x_{t+1}, \cdots ,x_{t+D-1},x_{t+D})$. This corresponds to indexing each $t$ to the $D$-dimensional vector $\textbf{s}(t)$.
The greater the value of $D$, the more information is incorporated into the vectors.

We should identify and count the number of occurrences of
all $D!$ permutations $\pi_j$ of length $D$ (with $j=1,2,...,D!)$.
The specific $j-th$ ordinal pattern associated to $\textbf{s}(t)$ is the permutation $\pi_j=(r_{0},r_{1},...,r_{D-1})_j$ of $(0,1,...,D-1)$ which guarantees that $x_{t+r_{0}} \leqslant x_{t+r_{1}} 
  \leqslant  \cdots  \leqslant x_{t+r_{(D-2)}} \leqslant x_{t+r_{(D-1)}} $. 
In order to get a unique result, we set $r_i < r_{i+1}$ if $x_{t+r_i} = x_{t+r_{i+1}}$.
In other words, each permutation $\pi_j$ is one of our possible symbols, and we have $D$! different symbols.
Therefore, the pertinent symbolic data is created by the following rules: (i) grouping the $D$ consecutive values of the time series points in the vector $\textbf{s}(t)$, (ii) indexing a symbol $\pi_j$ to the vector $\textbf{s}(t)$ by reordering the embedded data in ascending order using the permutation $\pi_j$.

Afterward, it is possible to quantify the diversity of the symbols in a scalar time series by counting how many times each one of the $D$! different permutations $\pi_j$ have been found in the dataset. Then, to calculate the PDF (for a specific $D$), we find
$P\equiv \{p_j;j=1,2,...,D! \}$, where $p_j$ is the probability to find the $j$-th symbol $\pi_j$ in our time series. This procedure is essential to a phase-space reconstruction with embedding dimension (pattern length) $D$. For practical purposes, BP suggested using $ 3\leqslant D \leqslant 7 $.

To have an example, choosing $D=3$, all six possible symbols (patterns) associated with the permutations $\pi_j$ are presented in Fig.~\ref{fig:symbol}(a). Considering the time series $X(t)=\{4,9,6,3,5,8,2,9,6\}$ as an example (see Fig.~\ref{fig:symbol}(b)), the first vector is $\textbf{s}(t=1)=(4,9,6)$, corresponding to the permutation $\pi_2=(0,2,1)$; the second vector is $\textbf{s}(t=2)=(9,6,3)$, corresponding to to the permutation $\pi_6=(2,1,0)$. Similarly, one can find the other five vectors $\textbf{s}(t)$ and their respective $\pi_j$. The corresponding non-normalized PDF is shown in Fig.~\ref{fig:symbol}(c).

It is also possible to associate a different PDF to the same time series $X(t)$, considering different values of time embedding $\tau$. This means that we can skip every $\tau-1$ points of the time series $X(t)$ in order to define the vector $\textbf{s}_{\tau}(t)$.
In our example this means that for $\tau=2$, the first vector is $\textbf{s}_{\tau=2}(t=1)=(4,6,5)$, corresponding to the permutation $\pi_2=(0,2,1)$; the second vector is $\textbf{s}_{\tau=2}(t=2)=(9,3,8)$, corresponding to the permutation $\pi_4=(2,0,1)$; the third vector is $\textbf{s}_{\tau=2}(t=3)=(6,5,2)$, corresponding to the permutation $\pi_6=(2,1,0)$; the fourth vector is $\textbf{s}_{\tau=2}(t=4)=(3,8,9)$, corresponding to the permutation $\pi_1=(0,1,2)$; and the last possible vector is $\textbf{s}_{\tau=2}(t=5)=(5,2,6)$, corresponding to the permutation $\pi_3=(1,0,2)$. 
For completness, for $\tau=3$ we have only 3 possible vectors: $\textbf{s}_{\tau=3}(t=1)=(4,3,2)$, $\textbf{s}_{\tau=3}(t=2)=(9,5,9)$, $\textbf{s}_{\tau=3}(t=3)=(6,8,6)$.
Therefore, with each time series
X (t), we can associate many PDFs, each one for a different
value of the time delay $\tau$. Unless otherwise stated here, we
evaluate the PDF for $\tau$ up to 20. Since the sampling rate of
our data is 250 Hz, $\tau=1,2,3$ represents a timescale of $4,8,16$~ms 

It is important to note that symbol sequences emerge directly from the time series without relying on model-based assumptions. Although this approach results in some loss of amplitude details from the original series, it preserves the temporal structure and provides information about the system’s temporal correlations. Lastly, the BP methodology only requires a very weak stationarity assumption: for $k \leq D$, the probability for $x_t \leq x_t+k$ should not depend on $t$.


\begin{figure*}
\centering
\begin{minipage}{5cm}
\begin{flushleft}(a)%
\end{flushleft}%
\centering
\includegraphics[width=0.98 \columnwidth,clip]{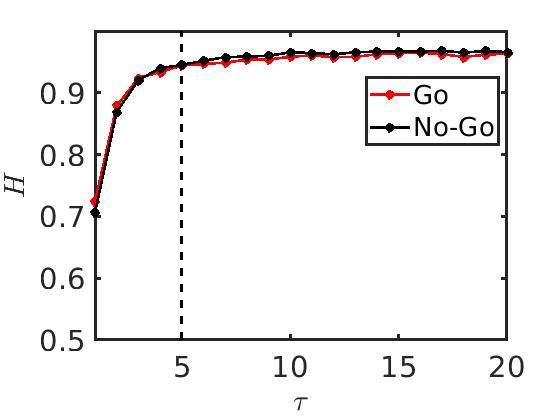}
\end{minipage}
\begin{minipage}{5cm}
\begin{flushleft}(b)%
\end{flushleft}%
\centering
\includegraphics[width=0.98\columnwidth,clip]{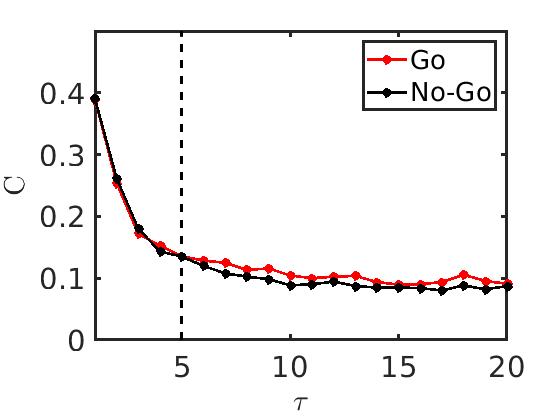}
\end{minipage}
\begin{minipage}{5cm}
\begin{flushleft}(c)%
\end{flushleft}%
\centering
\includegraphics[width=0.98\columnwidth,clip]{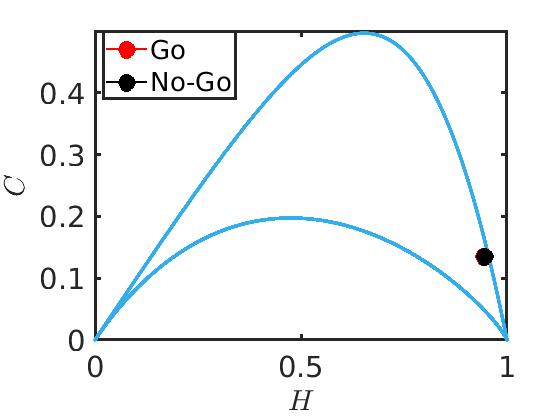}
\end{minipage}
\centering
\begin{minipage}{5cm}
\begin{flushleft}(d)%
\end{flushleft}%
\centering
\includegraphics[width=0.98 \columnwidth,clip]{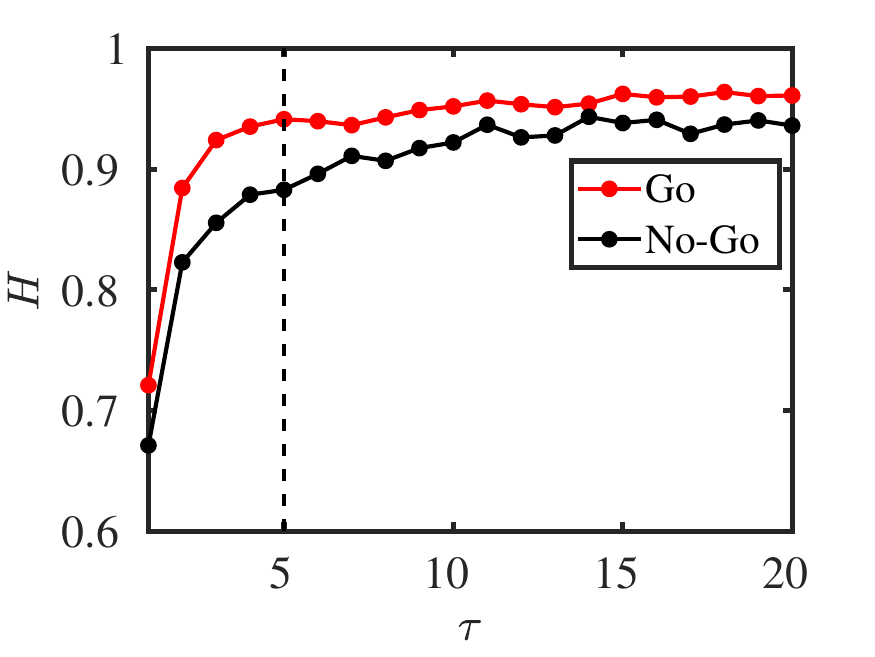}
\end{minipage}
\begin{minipage}{5cm}
\begin{flushleft}(e)%
\end{flushleft}%
\centering
\includegraphics[width=0.98\columnwidth,clip]{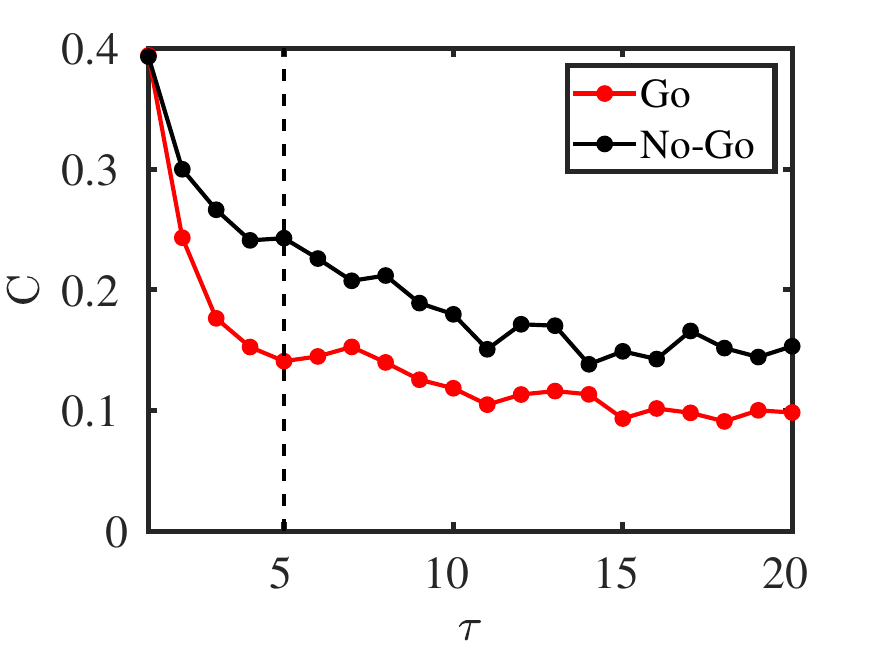}
\end{minipage}
\begin{minipage}{5cm}
\begin{flushleft}(f)%
\end{flushleft}%
\centering
\includegraphics[width=0.98\columnwidth,clip]{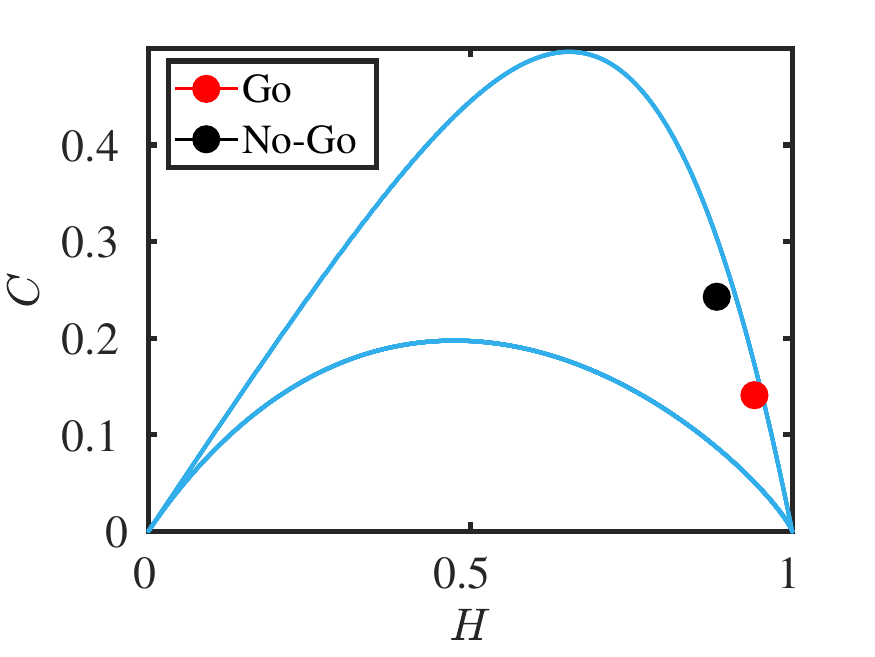}
\end{minipage}
\caption{
{\bf Entropy and Complexity can differentiate Go and NoGo trials in the central channel $C_z$ of an exemplar subject. } 
(a)-(c) Information theory indices calculated at time window starting at $t=300$~ms.
(a) Entropy for each trial type $H_\mathrm{Go}$ and $H_\mathrm{NoGo}$, and (b) Complexity $C_\mathrm{Go}$ and $C_\mathrm{NoGo}$ as a function of time embedding $\tau$ up to 20.
(c) Complexity-entropy ($HxC$) plane for $\tau=5$. (d)-(e) Similar to the upper panels but for a later time window starting at $t=1680$~ms. The indices can differentiate the trial type.
}
\label{fig:CxH_ch10}
\end{figure*}

\subsection{\label{Information theory quantifiers}{Information theory quantifiers}}

After calculating a probability distribution function (PDF) associated with a time series $X(t)$, any information-theory quantifier can be defined as a function of the PDF.
Using $P\equiv \{p_j;j=1,2,...,D! \}$  with $\sum_{j=1}^{N} p_j=1$ where $N=D!$ is the number of possible states of the system, one can calculate Shannon’s logarithmic information measure of a given time series using the following definition~\cite{Shannon49}:
\begin{equation}
\label{eq:S}
S[P] = -\sum_{j=1}^{N} p_j \ln(p_j).
\end{equation}
This function is equal to zero when we can correctly predict the outcome every time. 
By contrast, the entropy is maximized for the uniform distribution $P_e = \{p_j=1/N, \forall j=1,2, \dots , N\}$.
Then, the normalized Shannon entropy is defined by 
$H[P] = S[P] / S[P_e]$ ($0 \leq H \leq 1$).

For example, an ideal crystal is perfectly ordered, and a small amount of information is enough to describe the system. 
Thus, the entropy of such a crystal is minimal. At the opposite extreme, it would be the ideal gas, a completely disordered system, which would have a high value of entropy.
In an intermediate situation, one can find complex systems that cannot be fully characterized only by a randomness measure such as entropy. Thus, measures of statistical complexity are needed to gain a better understanding of time series representing such complex systems.
López-Ruiz et al.~\cite{lopez1995statistical} defined a complexity measure based on a disequilibrium function, which is related to the distance from an equiprobable PDF.

Here, we consider the MPR statistical complexity~\cite{Lamberti04}, 
based on the seminal notion of statistical complexity proposed by López-Ruiz et al.~\cite{lopez1995statistical}, defined through the product:
\begin{equation}
\label{eq:C}
C[P] = Q_J[P,P_e] \cdot H[P].
\end{equation}
The disequilibrium $Q_J[P,P_e]$ is defined in terms of the Jensen–Shannon divergence as:
\begin{equation}
\label{eq:Q}
Q_J[P,P_e] = Q_0 J[P,P_e],
\end{equation}
where
\begin{equation}
\label{eq:J}
J[P,P_e] = S\left[\frac{(P+P_e)}{2}\right] - \frac{S[P]}{2} - \frac{S[P_e]}{2},
\end{equation}
and $Q_0$ is a normalization constant ($0 \leq Q_J \leq 1$), equal to the inverse of the maximum possible value of $J[P, P_e]$.
This maximum value is obtained when one of the components of $P$, say $p_m$, is equal to 1, and the remaining $p_j$ are equal to zero.

The Jensen–Shannon divergence $J[P,P_e]$ is used to quantify the difference between two probability distributions: $P$ and $P_e$, respectively, the one associated with the system of interest and the uniform distribution. It is especially advantageous to compare the symbolic composition between different sequences~\cite{Grosse02}. 
It has been shown that, for a given value of normalized entropy $H$, the complexity $C$ can vary between a well-defined minimum $C_\mathrm{min}$ and a maximum $C_\mathrm{max}$ value, which restricts the possible occupied region in the complexity-entropy plane~\cite{Martin06}. 

 We can define a distance between entropy and complexity for two conditions $i$ and $j$ as the simple Euclidean distance:
\begin{equation}
\label{eq:distance}
    D_{i,j}=\sqrt{ (H_{i} - H_{j})^2 + (C_{i}-C_{j})^2}.
\end{equation}





\begin{figure*}
\centering
\begin{minipage}{5cm}
\begin{flushleft}(a)%
\end{flushleft}%
\centering
\includegraphics[width=0.98\columnwidth,clip]{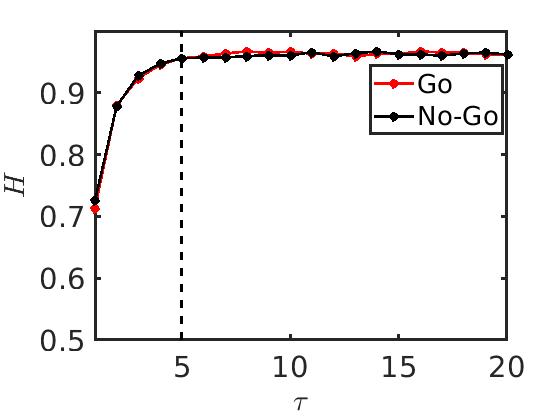}
\end{minipage}
\begin{minipage}{5cm}
\begin{flushleft}(b)%
\end{flushleft}%
\centering
\includegraphics[width=0.98\columnwidth,clip]{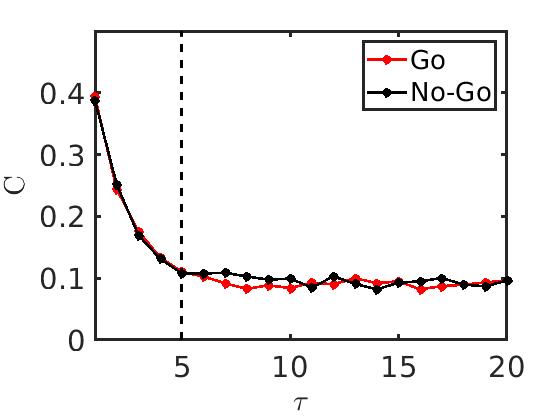}
\end{minipage}
\begin{minipage}{5cm}
\begin{flushleft}(c)%
\end{flushleft}%
\centering
\includegraphics[width=0.98\columnwidth,clip]{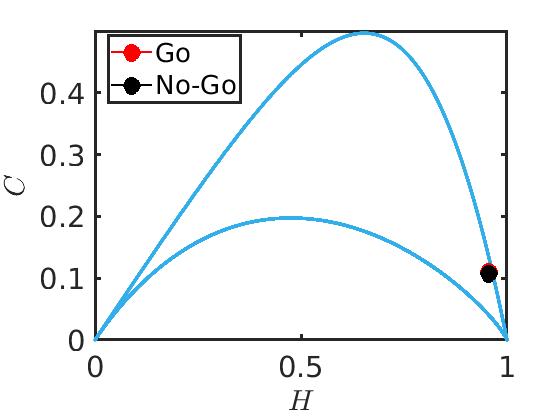}
\end{minipage}
\begin{minipage}{5cm}
\begin{flushleft}(d)%
\end{flushleft}%
\centering
\includegraphics[width=0.98\columnwidth,clip]{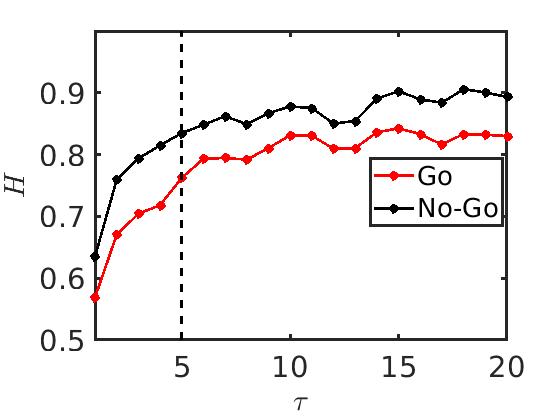}
\end{minipage}
\begin{minipage}{5cm}
\begin{flushleft}(e)%
\end{flushleft}%
\centering
\includegraphics[width=0.98\columnwidth,clip]{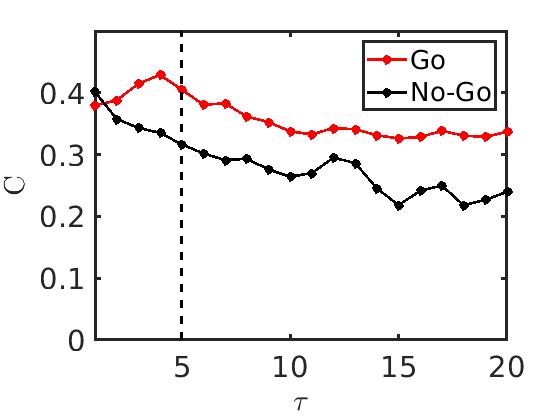}
\end{minipage}
\begin{minipage}{5cm}
\begin{flushleft}(f)%
\end{flushleft}%
\centering
\includegraphics[width=0.98\columnwidth,clip]{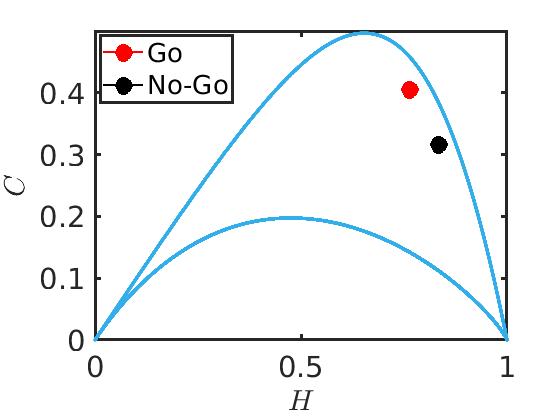}
\end{minipage}
\caption{
\textbf{Entropy and Complexity can distinguish trial types in an occipital channel $O_2$ of an exemplar subject.} 
(a)-(c) Information theory indices calculated at time window starting at $t=300$~ms.
(a) Entropy for each trial type $H_\mathrm{Go}$ and $H_\mathrm{NoGo}$, and (b) Complexity $C_\mathrm{Go}$ and $C_\mathrm{NoGo}$ as a function of time embedding $\tau$ up to 20.
(c) Complexity-entropy ($HxC$) plane for $\tau=5$. (d)-(e) Similar to the upper panels but for a later time window starting at $t=1540$~ms. The indices can differentiate the trial type.
}
\label{fig:CxH_ch19}
\end{figure*}


\begin{figure}
\centering
\begin{minipage}{8cm}
\begin{flushleft}(a)%
\end{flushleft}%
\includegraphics[width=0.9\columnwidth,clip]{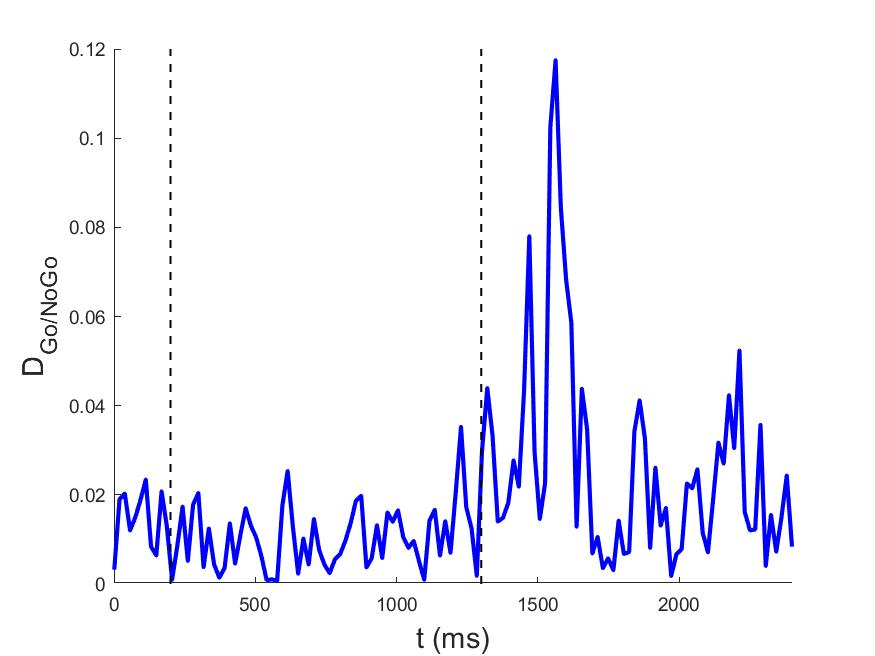}
\end{minipage}
\begin{minipage}{8cm}
\begin{flushleft}(b)%
\end{flushleft}%
\centering
\includegraphics[width=0.9\columnwidth,clip]{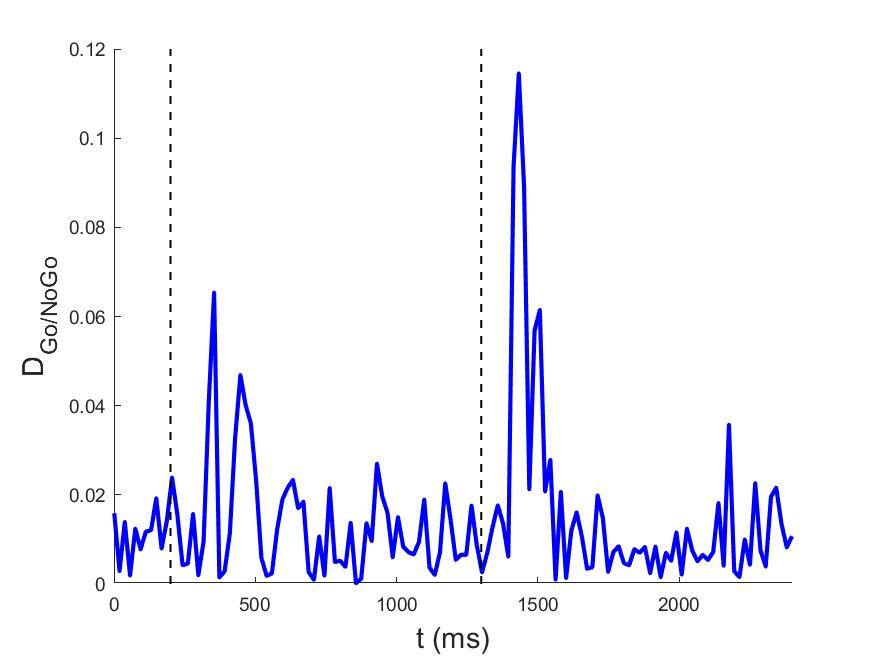}
\end{minipage}
\begin{minipage}{8cm}
\begin{flushleft}(c)%
\end{flushleft}%
\centering
\includegraphics[width=0.9\columnwidth,clip]{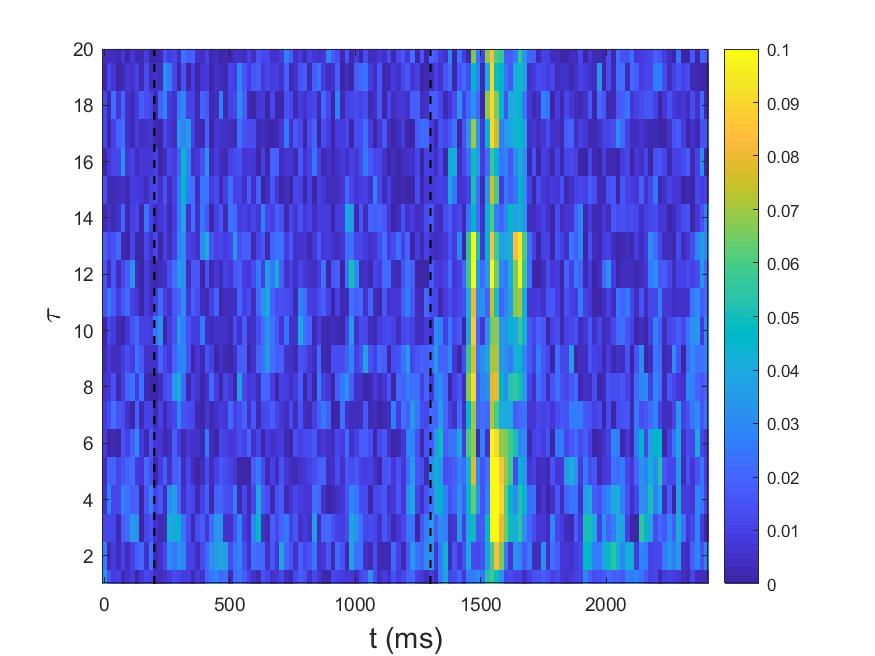}
\end{minipage}
\begin{minipage}{8cm}
\begin{flushleft}(d)%
\end{flushleft}%
\centering
\includegraphics[width=0.9\columnwidth,clip]{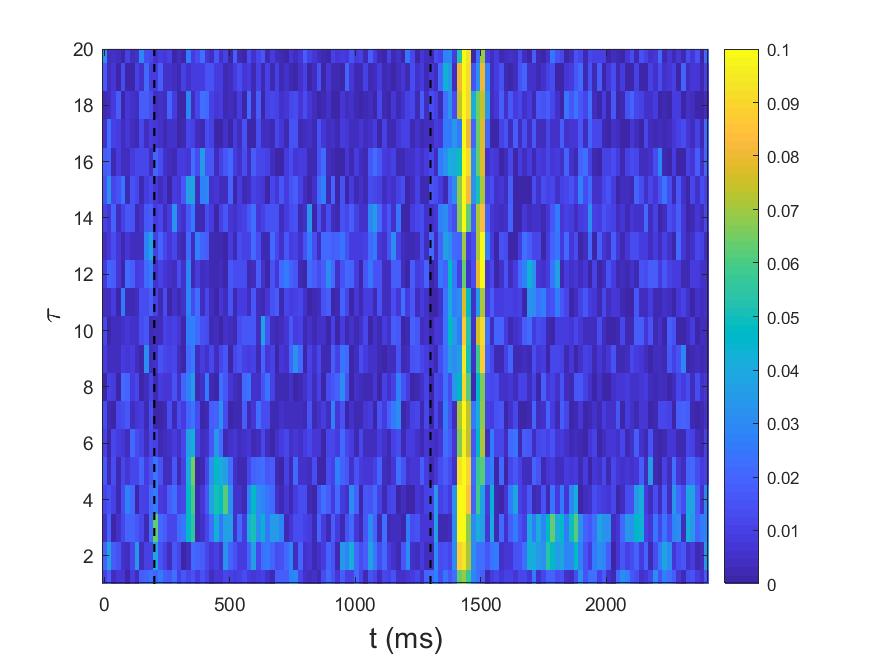}
\end{minipage}
\caption{
\textbf{Representation of the Euclidean distance between the Go and NoGo conditions ($D_\mathrm{Go/NoGo}$) along the entire trial for channels $C_z$ and $O_2$ (first and second columns, respectively) of an exemplar subject.} (a) and (b) $D_\mathrm{Go/NoGo}$ as a function of time for fixed $\tau=5$.
(c) and (d) Heatmap representing $D_\mathrm{Go/NoGo}$ in the 2D diagram of $\tau$ versus time along the trial. 
} 
\label{fig:distance_ch10_ch19}
\end{figure}



\section{\label{results}Results}


\subsection{Characterizing Go and NoGo trials at the individual level}

We employ information theory quantifiers $H$ and $C$ as a function of the time embedding $\tau$ to study response-specific processes in EEG signals. We aim to characterize cortical information processes during a visuomotor experimental paradigm related to a Go/NoGo task.  
In each trial, two visual stimuli are presented in sequence: the first one at $t_a=300$~ms and the second one at $t_b=1400$~ms. Depending on the combination of the two stimuli, the participants are supposed to press a button (Go trials), or the participants should suppress the action of pressing the button (NoGo trials, see Sec.~\ref{subsec:task} for more details).
We have separately analyzed the EEG time series of all 100 Go-response trials and all 100 NoGo-response trials for the 19 channels of 11 subjects.
For each one of those electrodes and volunteers, we use the concatenated time series of all Go trials to calculate $H_\mathrm{Go}$ and $C_\mathrm{Go}$ in each $100$~ms window starting every $20$~ms along the trial duration. Then, we repeat the same analysis for all NoGo trials and obtain $H_\mathrm{NoGo}$ and $C_\mathrm{NoGo}$ as a function of the time embedding $\tau$ for each time window (see Sec.~\ref{subsec:timeseries}  for more details).

One should expect no statistical differences between the two trial types in any time window before the second image appears on the screen ($t leq t_b=1400$~ms). In fact, this is what happens. After the second image, it takes a while for the statistical properties of the signals to start differing, which is related to the time scales relative to cognitive processes.

\subsubsection{Distinguishing trial types at an exemplar channel}

Initially, we analyze the indices for the central electrode $C_z$ of one specific subject in early time windows. In particular, but with no loss of generality, we chose to show the results at $t=600$~ms in Figs.~\ref{fig:CxH_ch10}(a),(b),(c). Then we repeat the analysis in a later time window, after the second cue ($t \geq 1400$~ms).   
In Figs.~\ref{fig:CxH_ch10}(a) and (b) we show the $H_\mathrm{Go}$ and $C_\mathrm{Go}$ in red and $H_\mathrm{NoGo}$ and $C_\mathrm{NoGo}$ in black as a function of the embedding time $\tau$ at $t=600$~ms. Neither $H$ nor $C$ can distinguish the Go/NoGo conditions at this time window. This means that $H_\mathrm{Go}\approx H_\mathrm{NoGo}$ and $C_\mathrm{Go}\approx C_\mathrm{NoGo}$ for every $\tau$. This is merely a consistency check, as the subject has no information about the trial type at this moment.  

In Figs.~\ref{fig:CxH_ch10}(d) and (e) we show the same indexes in a different time window: $t=1680$~ms. At this later time, both $H$ and $C$ can distinguish the trial type:  
$H_\mathrm{Go}\neq H_\mathrm{NoGo}$ and $C_\mathrm{Go}\neq C_\mathrm{NoGo}$ for many values of $\tau$. 
In particular, at this later time window, the entropy of Go trials $H_\mathrm{Go}$ remains similar to the values at $t=600$~ms, but the entropy of the NoGo trials $H_\mathrm{NoGo}$ decreases: $H_\mathrm{NoGo}(t=1680$~ms$)<H_\mathrm{NoGo}(t=300$~ms$)$. 

Comparing Figs.~\ref{fig:CxH_ch10}(b) and (e), we can see what happens for the complexity: $C_\mathrm{NoGo}$ at $t=1680$~ms is larger than $C_\mathrm{NoGo}$ at $t=600$~ms. This indicates that the indices for Go and NoGo are different after the time point at which the subject already knows if it is a Go or a NoGo trial. This means that the statistical properties of the activity in this channel differ for Go and NoGo trials.

In Fig.~\ref{fig:CxH_ch10}(c) and (f), we show that for $\tau=5$ (which represents a time scale of 20~ms) the two conditions can be separated in the $HxC$ plane at $t=1680$~ms, but not at $t=600$~ms. This is good evidence that we can use the $CxH$ plane to separate response-specific trial types in EEG data. Results are qualitatively similar for many values of $\tau$, including $2<\tau<10$.

It is worth mentioning that we can visualize the decrease of entropy, together with the increase of complexity for NoGo trials in the CxH plane, noticing that the black circle moves away from the region of the plane which represents the noisiest signals: the corner $H=1$ and $C=0$ (compare Figs.~\ref{fig:CxH_ch10}(c) and (f)). This means that the brain activity captured by this channel at $t=1680$~ms for the NoGo condition is less noisy (less uncorrelated) than the activity for Go trials.

At this point, we should emphasize that in Fig.~\ref{fig:CxH_ch10}(C) and (f) we show the $CxH$ plane for $\tau=5$, channel $C_z$, and one subject, in two different time windows $t=600$~ms and $t=1680$~ms. However, the separation between the two trial types has been consistently verified for other values of $\tau$, other time windows after the second cue, other channels, and other subjects, as we will show in the following figures. 


\vspace{-0.15cm}
\begin{figure*}[h]
\centering
\begin{minipage}{0.23\linewidth}
\vspace{-0.2cm}
\begin{flushleft}(a)%
\end{flushleft}%
\centering
\includegraphics[width=0.98 \columnwidth,clip]{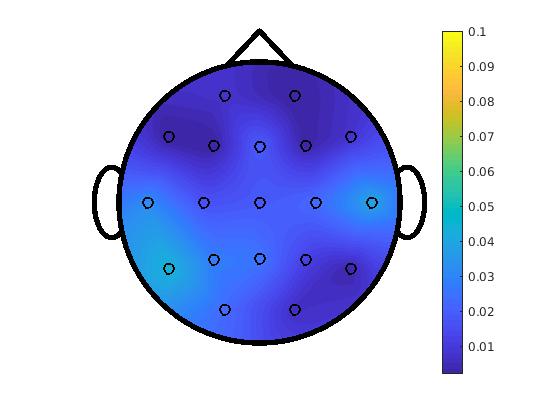}
\end{minipage}
\begin{minipage}{0.23\linewidth}
\vspace{-0.2cm}
\begin{flushleft}(b)%
\end{flushleft}%
\centering
\includegraphics[width=0.98\columnwidth,clip]{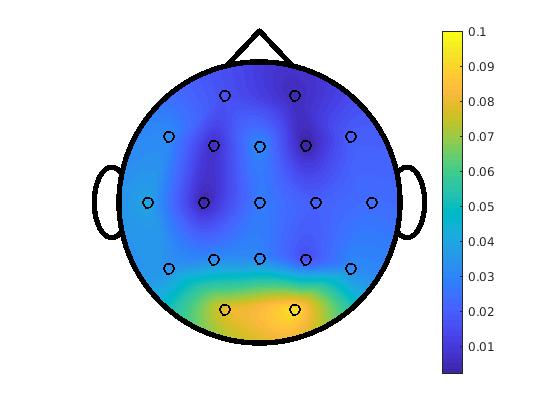}
\end{minipage}
\begin{minipage}{0.23\linewidth}
\vspace{-0.2cm}
\begin{flushleft}(c)%
\end{flushleft}%
\centering
\includegraphics[width=0.98 \columnwidth,clip]{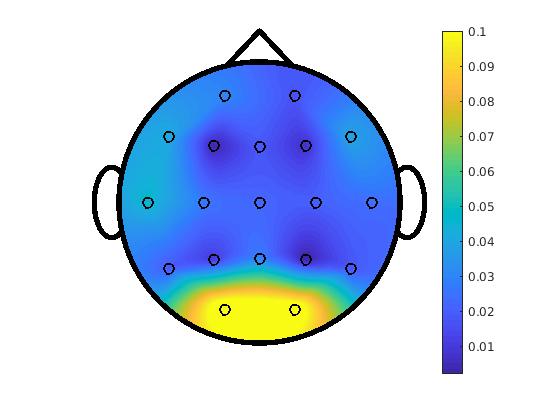}
\end{minipage}
\begin{minipage}{0.23\linewidth}
\vspace{-0.2cm}
\begin{flushleft}(d)%
\end{flushleft}%
\centering
\includegraphics[width=0.98 \columnwidth,clip]{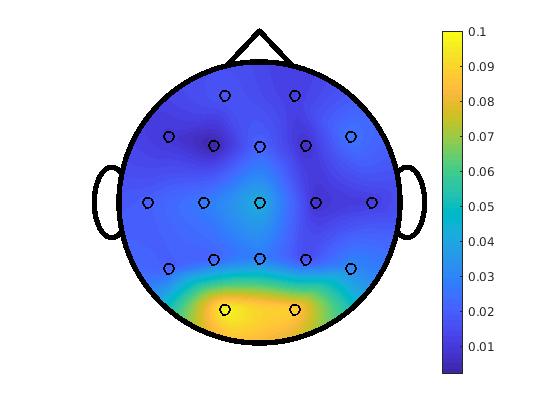}
\end{minipage}
\centering
\begin{minipage}{0.23\linewidth}
\vspace{-0.2cm}
\begin{flushleft}(e)%
\end{flushleft}%
\centering
\includegraphics[width=0.98 \columnwidth,clip]{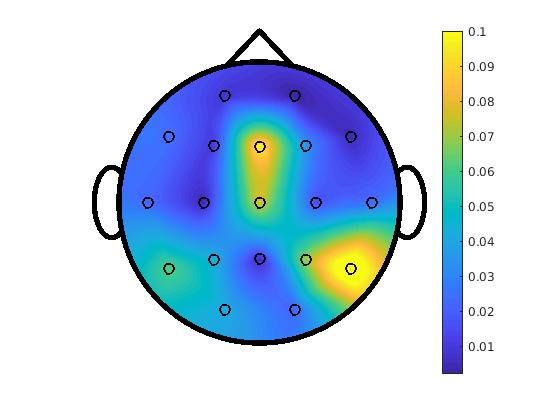}
\end{minipage}
\begin{minipage}{0.23\linewidth}
\vspace{-0.2cm}
\begin{flushleft}(f)%
\end{flushleft}%
\centering
\includegraphics[width=0.98 \columnwidth,clip]{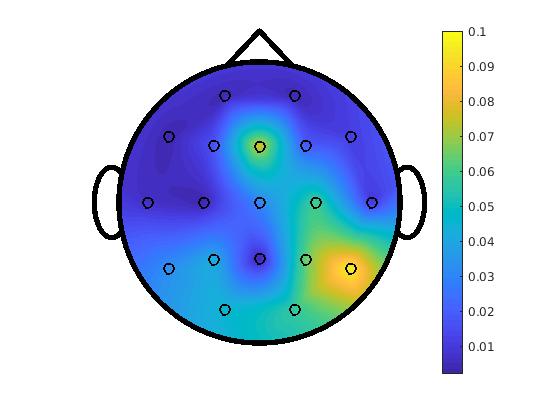}
\end{minipage}
\begin{minipage}{0.23\linewidth}
\vspace{-0.2cm}
\begin{flushleft}(g)%
\end{flushleft}%
\centering
\includegraphics[width=0.98 \columnwidth,clip]{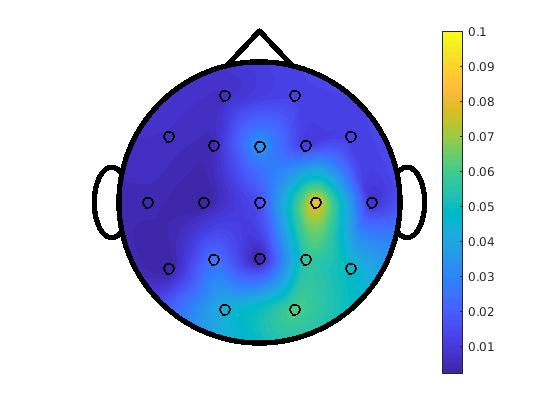}
\end{minipage}
\begin{minipage}{0.23\linewidth}
\vspace{-0.2cm}
\begin{flushleft}(h)%
\end{flushleft}%
\centering
\includegraphics[width=0.98 \columnwidth,clip]{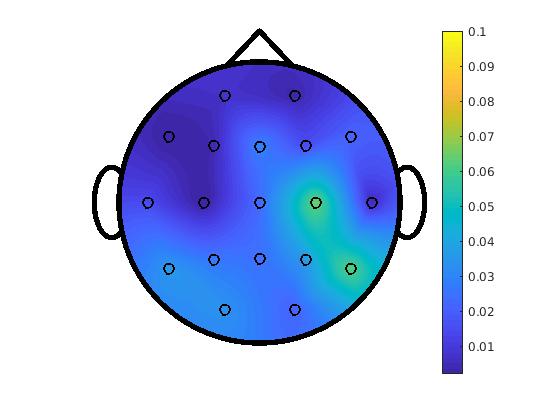}
\end{minipage}
\centering
\begin{minipage}{0.23\linewidth}
\vspace{-0.2cm}
\begin{flushleft}(i)%
\end{flushleft}%
\centering
\includegraphics[width=0.98 \columnwidth,clip]{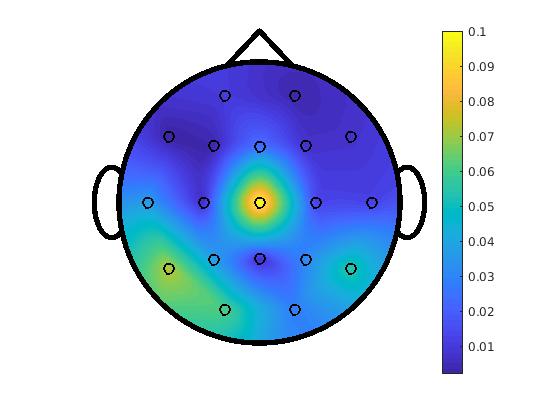}
\end{minipage}
\begin{minipage}{0.23\linewidth}
\vspace{-0.2cm}
\begin{flushleft}(j)%
\end{flushleft}%
\centering
\includegraphics[width=0.98 \columnwidth,clip]{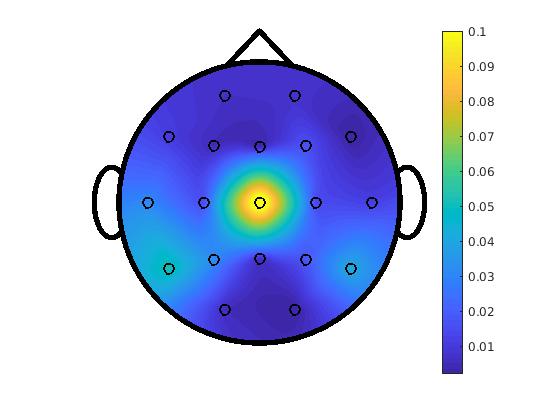}
\end{minipage}
\begin{minipage}{0.23\linewidth}
\vspace{-0.2cm}
\begin{flushleft}(k)%
\end{flushleft}%
\centering
\includegraphics[width=0.98 \columnwidth,clip]{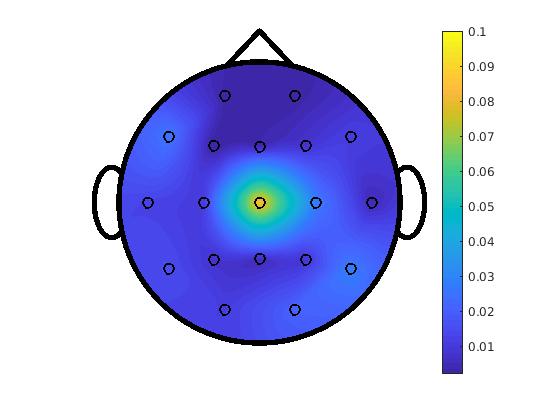}
\end{minipage}
\begin{minipage}{0.23\linewidth}
\vspace{-0.2cm}
\begin{flushleft}(l)%
\end{flushleft}%
\centering
\includegraphics[width=0.98 \columnwidth,clip]{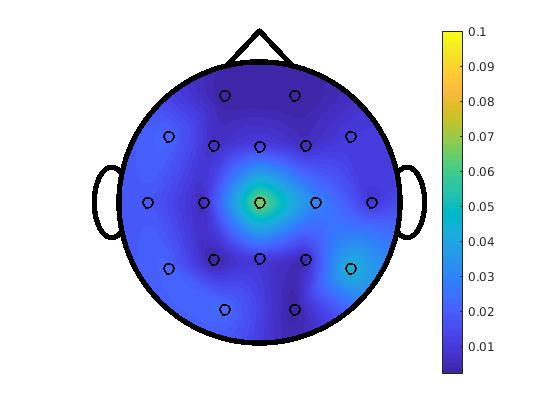}
\end{minipage}
\caption{\label{fig:heatmapheadonesubject}
{\bf Topographic brain heatmap: the Euclidean distance between the two conditions evaluated in 12 intervals. }
(a) $t=1500$~ms (b) $t=1520$~ms, (c) $t=1540$~ms, up to (l) $t=1720$~ms.
}
\end{figure*}




\begin{figure}
\centering
\begin{minipage}{8cm}
\begin{flushleft}(a)%
\end{flushleft}%
\centering
\includegraphics[width=0.9\columnwidth,clip]{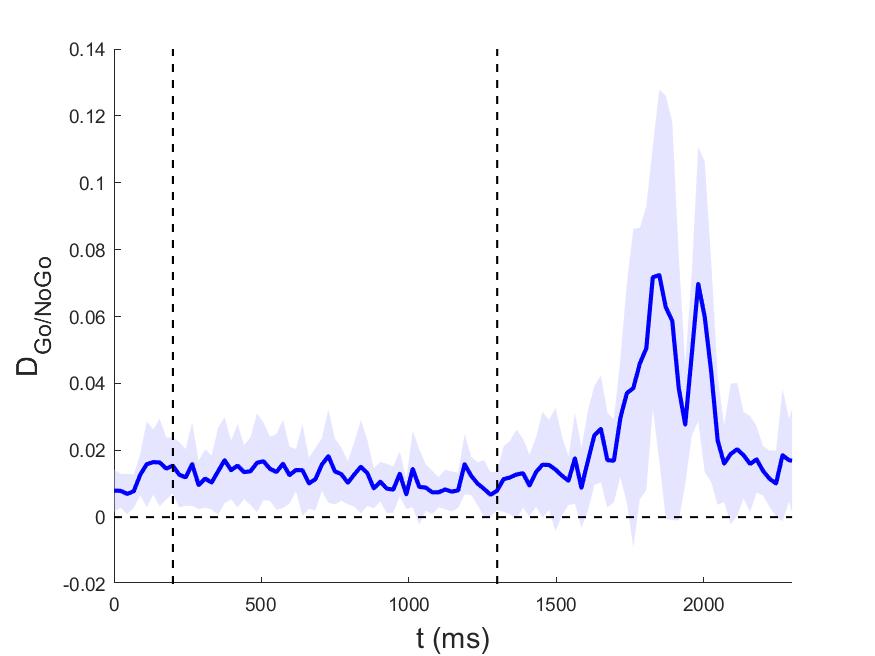}
\end{minipage}
\begin{minipage}{8cm}
\begin{flushleft}(b)%
\end{flushleft}%
\centering
\includegraphics[width=0.9\columnwidth,clip]{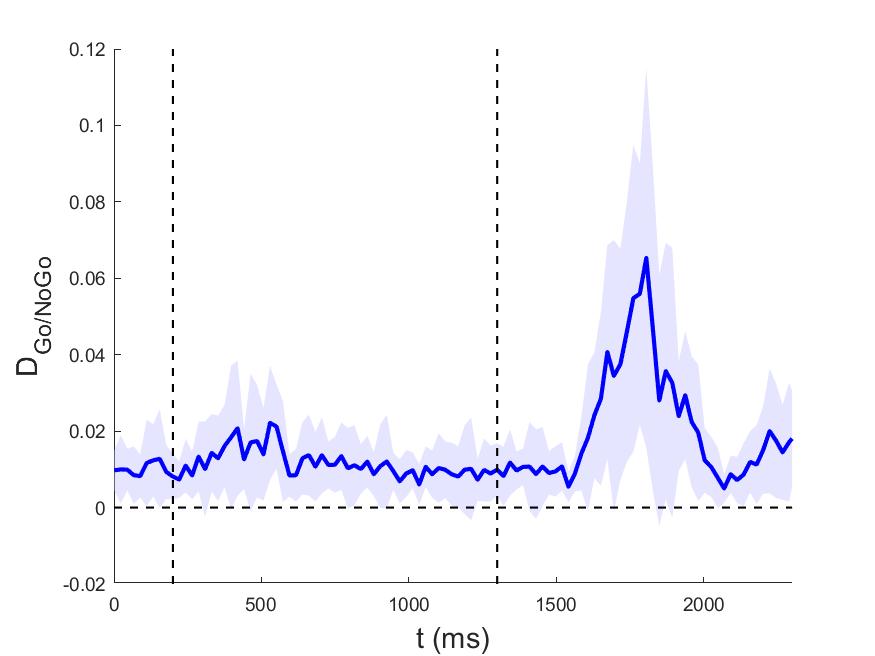}
\end{minipage}
\begin{minipage}{8cm}
\begin{flushleft}(c)%
\end{flushleft}%
\centering
\includegraphics[width=0.98\columnwidth,clip]{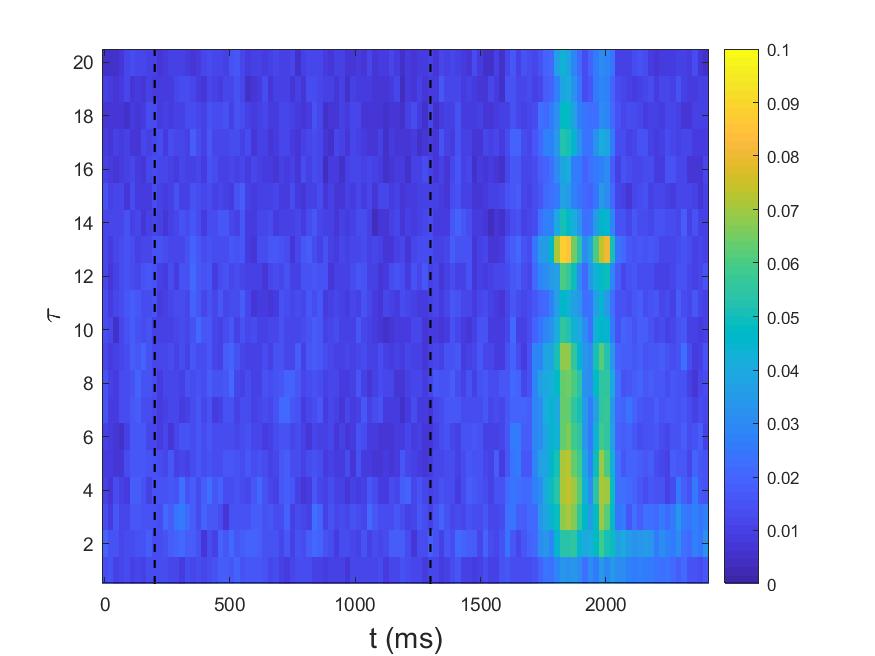}
\end{minipage}
\begin{minipage}{8cm}
\begin{flushleft}(d)%
\end{flushleft}%
\centering
\includegraphics[width=0.98\columnwidth,clip]{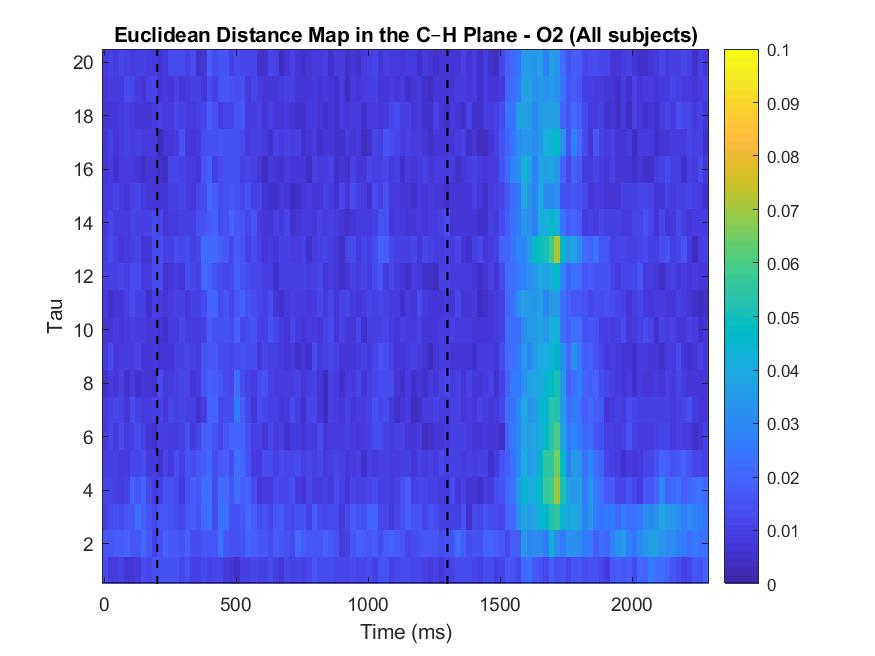}
\end{minipage}
\caption{
{\bf The Euclidean distance in the $CxH$ plane between the Go and NoGo conditions $D_\mathrm{Go/NoGo}$ can distinguish the trial types at the group level.} 
(a) and (b) Average value of $D_\mathrm{Go/NoGo}$ for all subjects as a function of time for fixed $\tau=5$.
(c) and
(d) Heatmap representing $D_\mathrm{Go/NoGo}$ in the 2D diagram of $\tau$ versus time along the trial. 
The first column represents channels $C_z$ and the second, channels $O_2$.
}
\label{fig:group}
\end{figure}

\subsubsection{Results are robust in different channels, time windows, and time scales}

Differences in the indices, and consequently the separation between two conditions in the $CxH$ plane, can also be verified in other channels of the same subject. For example, in Fig.~\ref{fig:CxH_ch19}, we show the same analysis as in Fig.~\ref{fig:CxH_ch10} but for the occipital electrode $O_2$. The time window in which the two conditions are more separated can be different from channel to channel. For $O_2$ it occurs at the time window associated with the interval $t=1540$~ms to $t=1640$~ms (see 
Figs.~\ref{fig:CxH_ch19}(d), (e), (f)). The indices for an early time window, which does not present any distinction between trial types, are shown in
Figs.~\ref{fig:CxH_ch19}(a), (b), (c) calculated at $t=600$~ms.

 Interestingly, the behavior of entropy and complexity after the second image differs when comparing the central channel $C_z$ and the occipital channel $O_2$. For $O_2$,
both the Go and NoGo trials move away from the noisiest region of the plane after the second cue. Moreover, the entropy of the Go trials decreases more than that of the NoGo trials. This behavior can be verified by comparing the relative positions of black and red dots in figures~\ref{fig:CxH_ch10}(f) and Fig.~\ref{fig:CxH_ch19}(f).


We then calculated the Euclidean distance as a function of the time embedding $\tau$ for each time window. 
In Fig~\ref{fig:distance_ch10_ch19}(a) and (b), we see the distance between the two conditions $D_\mathrm{Go/NoGo}$ as a function of the time interval along the trial for $\tau=5$. The time window in which the distance is maximal is marked in blue ($t=1680$~ms for $C_z$ (Fig~\ref{fig:distance_ch10_ch19}(b)) and $t=1540$~ms for $O_2$ (Fig~\ref{fig:distance_ch10_ch19}(d)).  
Dashed black lines indicate the time at which the images appear ($t_a=300$~ms and $t_b=1400$~ms). No significant values of distance should be verified before $t_b$.

In Fig~\ref{fig:distance_ch10_ch19}(c) and (d), we show the heatmap of the distance between the two conditions $D_\mathrm{Go/NoGo}$.
The heatmap indicates that many values of $\tau$ can be used to distinguish between the trial types. 
Therefore, the Euclidean distance is a suitable proxy for distinguishing Go and NoGo trials across different brain regions and time windows. 

To visualize at a glance what is happening in the entire brain, Fig.~\ref{fig:heatmapheadonesubject} shows a brain heatmap of the distance for different time windows, ranging from $t=1500$~ms to $t=1720$~ms.
For example, the separation between trial types begins at $O_1$ and $O_2$ moves to more lateral and central areas, and can ultimately be verified at $C_z$.  
In Fig.~\ref{fig:heatmapheadonesubject}(c) we show the same interval as in Fig.~\ref{fig:CxH_ch19} ($t=1540$~ms). 
In Fig.~\ref{fig:heatmapheadonesubject}(j) we show the same interval as in Fig.~\ref{fig:CxH_ch10} ($t=1680$~ms).

One could ask what it means that only specific channels show differences between the two conditions. We suggest that not all of the brain is involved in computing differences between trials. Alternatively, at least these channels are powerfully capturing the cognitive differences related to Go and NoGo conditions.

\subsection{Distinguishing trial types at the group level}

At this point, we can analyze the statistical properties of all subjects together and for each channel separately. Despite the individual variability regarding the $\tau$ that maximizes the distance $D_\mathrm{Go/NoGo}$ and the time windows in which the trial types can start to be differentiated, results are robust on average.

In Fig~\ref{fig:group}(a) and (b), we see the average distance between the two conditions $D_\mathrm{Go/NoGo}$ as a function of the time interval along the trial for $\tau=5$. 
Dashed black lines indicate the time at which the images appear ($t_a=300$~ms and $t_b=1400$~ms). No significant values of distance should be verified before $t_b$.

In Fig~\ref{fig:group}(c) and (d), we show the heatmap of the distance between the two conditions $D_\mathrm{Go/NoGo}$.
The heatmap indicates that many values of $\tau$ can be used to distinguish between the trial types. 

\begin{figure}
\centering
 \begin{minipage}{19cm}
\hspace{-2,0cm}
\includegraphics[width=1.1\columnwidth,clip]{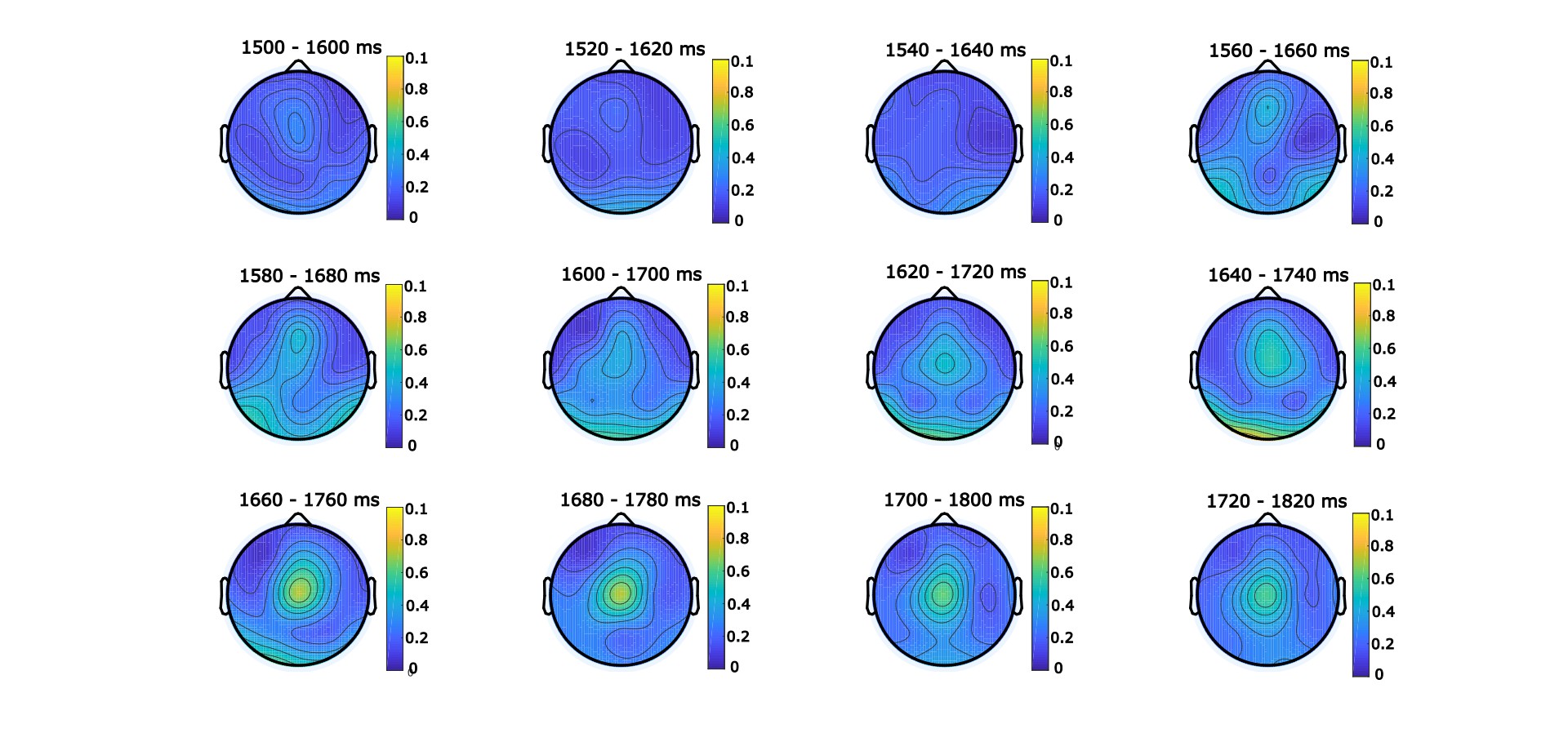}
\end{minipage}
\caption{
{\bf Topographic brain heatmap: the group average of the Euclidean distance between the two conditions systematically evaluated in 12 intervals. }
Starting from $t=1500$~ms, then $t=1520$~ms, and then showing a new interval every $20$~ms until $t=1720$~ms.}
\label{fig:groupheadsdistance}
\end{figure}

In Fig.~\ref{fig:groupheadsdistance}, we show 
in a brain heatmap the distance for different time windows from $t=1500$~ms to $t=1720$~ms.
Similarly to what happens at the individual level, the separation between trial types begins at $O_1$ and $O_2$ and moves to more central areas, reaching $C_z$.

\section{\label{conclusion}Conclusion}

In conclusion, we have demonstrated that information theory quantifiers
\cite{Shannon49,lopez1995statistical,Lamberti04, Rosso07, Zunino12, xiong2020complexity}, such as Shannon entropy, MPR statistical complexity, and the multiscale complexity-entropy plane, are effective tools for characterizing information processing in human EEG signals recorded during a visuomotor task.

Using EEG data from 11 healthy subjects~\cite{carlos2020anticipated}, we were able to distinguish between Go and NoGo trials and identify time windows and electrodes that are more relevant to these processes. Additionally, we identified cortical regions that exhibit more pronounced differences between the two trial types and described trial-type-specific dynamics across various time intervals.

The analysis was shown to be applicable at both the individual and group levels. These findings contribute to the investigation of stimulus- and response-specific brain activity.
Unlike certain classification approaches based on machine learning, the present method enables the interpretation of signals in terms of their physical properties, allowing for comparison with ordered and disordered states.

This framework may be extended to explore differences in more subjective tasks, such as memory \cite{bavassi2019retrieval}, as well as to cognitive tasks involving other type of recordings, such as MEG~\cite{Kosem16,michalareas2016alpha} and intracranial data~\cite{pinheiro2023direct,pinheiro2024spatiotemporal}. It also holds potential for distinguishing between clinical and control populations~\cite{echegoyen2020permutation}, and for characterizing brain states under conditions such as anesthesia and sleep.


\printcredits

\section*{Acknowledgment}
The authors thank CNPq (Grants No. 402359/2022-4
and No. 314092/2021-8), FAPEAL (Grant No.
APQ2022021000015), UFAL, CAPES, and LOREAL-UNESCO-ABC For Women In Science (Para Mulheres na
Ci\^encia) for financial support. This study was financed in part by the Coordenação de Aperfeiçoamento de Pessoal de Nível Superior - Brasil (CAPES) - Finance Code 001.

\bibliographystyle{elsarticle-num}  

\bibliography{matias}

\begin{thebibliography}{10}
\expandafter\ifx\csname url\endcsname\relax
  \def\url#1{\texttt{#1}}\fi
\expandafter\ifx\csname urlprefix\endcsname\relax\def\urlprefix{URL }\fi
\expandafter\ifx\csname href\endcsname\relax
  \def\href#1#2{#2} \def\path#1{#1}\fi

\bibitem{Ledberg07}
A.~Ledberg, S.~L. Bressler, M.~Ding, R.~Coppola, R.~Nakamura, Large-scale
  visuomotor integration in the cerebral cortex, Cerebral cortex 17~(1) (2007)
  44--62.

\bibitem{Bandt02}
C.~Bandt, B.~Pompe, Permutation entropy: a natural complexity measure for time
  series, Physical review letters 88~(17) (2002) 174102.

\bibitem{Shannon49}
C.~Shannon, W.~Weaver, The mathematical theory of communication, Champaign, IL:
  University of Illinois Press, 1949.

\bibitem{lopez1995statistical}
R.~Lopez-Ruiz, H.~L. Mancini, X.~Calbet, A statistical measure of complexity,
  Physics letters A 209~(5-6) (1995) 321--326.

\bibitem{Lamberti04}
P.~W. Lamberti, M.~T. Mart\'{\i}n, A.~Plastino, O.~A. Rosso, Intensive entropic
  non-triviality measure, Physica A: Statistical Mechanics and its Applications
  334 (2004) 119--131.

\bibitem{Rosso07}
O.~A. Rosso, H.~A. Larrondo, M.~T. Mart\'{\i}n, A.~Plastino, M.~Fuentes,
  Distinguishing noise from chaos, Physical Review Letters 99~(15) (2007)
  154102.

\bibitem{Zunino12}
L.~Zunino, M.~C. Soriano, O.~A. Rosso, Distinguishing chaotic and stochastic
  dynamics from time series by using a multiscale symbolic approach, Physical
  Review E 86 (2012) 046210.

\bibitem{xiong2020complexity}
H.~Xiong, P.~Shang, J.~He, Y.~Zhang, Nonlinear Dynamics 100 (2020) 1673--1687.

\bibitem{harmony2009time}
T.~Harmony, A.~Alba, J.~L. Marroqu{\'\i}n, B.~Gonz{\'a}lez-Frankenberger,
  Time-frequency-topographic analysis of induced power and synchrony of eeg
  signals during a go/no-go task, International Journal of Psychophysiology
  71~(1) (2009) 9--16.

\bibitem{Bressler93}
S.~L. Bressler, R.~Coppola, R.~Nakamura, Episodic multiregional cortical
  coherence at multiple frequencies during visual task performance, Nature
  366~(6451) (1993) 153--156.

\bibitem{Liang00}
H.~Liang, M.~Ding, R.~Nakamura, S.~L. Bressler, Causal influences in primate
  cerebral cortex during visual pattern discrimination, Neuroreport 11~(13)
  (2000) 2875--2880.

\bibitem{Brovelli04}
A.~Brovelli, M.~Ding, A.~Ledberg, Y.~Chen, R.~Nakamura, S.~L. Bressler, Beta
  oscillations in a large-scale sensorimotor cortical network: Directional
  influences revealed by {G}ranger causality, Proc. Natl. Acad. Sci. {USA}
  101~(26) (2004) 9849--9854.

\bibitem{Salazar12}
R.~F. Salazar, N.~M. Dotson, S.~L. Bressler, C.~M. Gray, Content-specific
  fronto-parietal synchronization during visual working memory, Science 338
  (2012) 1097--1100.

\bibitem{Dotson14}
N.~M. Dotson, R.~F. Salazar, C.~M. Gray, Frontoparietal correlation dynamics
  reveal interplay between integration and segregation during visual working
  memory, The Journal of Neuroscience 34~(41) (2014) 13600--13613.

\bibitem{deLucas21}
H.~B. de~Lucas, S.~L. Bressler, F.~S. Matias, O.~A. Rosso, A symbolic
  information approach to characterize response-related differences in cortical
  activity during a go/no-go task, Nonlinear Dynamics 104~(4) (2021)
  4401--4411.

\bibitem{da2024symbolic}
{\'I}.~R. S.~C. Da~Paz, P.~F. Silva, H.~B. de~Lucas, S.~H. Lira, O.~A. Rosso,
  F.~S. Matias, Symbolic information approach applied to human intracranial
  data to characterize and distinguish different congnitive processes, Physical
  Review E 110~(2) (2024) 024403.

\bibitem{Montani15}
F.~Montani, O.~A. Rosso, F.~S. Matias, S.~L. Bressler, C.~R. Mirasso, A
  symbolic information approach to determine anticipated and delayed
  synchronization in neuronal circuit models, Phil. Trans. R. Soc. A 373~(2056)
  (2015) 20150110.

\bibitem{lotfi2021statistical}
N.~Lotfi, T.~Feliciano, L.~A. Aguiar, T.~P.~L. Silva, T.~T. Carvalho, O.~A.
  Rosso, M.~Copelli, F.~S. Matias, P.~V. Carelli, Statistical complexity is
  maximized close to criticality in cortical dynamics, Physical Review E
  103~(1) (2021) 012415.

\bibitem{rosso2006eeg}
O.~Rosso, M.~Martin, A.~Figliola, K.~Keller, A.~Plastino, Eeg analysis using
  wavelet-based information tools, Journal of neuroscience methods 153~(2)
  (2006) 163--182.

\bibitem{echegoyen2020permutation}
I.~Echegoyen, D.~L{\'o}pez-Sanz, J.~H. Mart{\'\i}nez, F.~Maest{\'u}, J.~M.
  Buld{\'u}, Permutation entropy and statistical complexity in mild cognitive
  impairment and alzheimer’s disease: An analysis based on frequency bands,
  Entropy 22~(1) (2020) 116.

\bibitem{jungmann2024state}
R.~M. Jungmann, T.~Feliciano, L.~A. Aguiar, C.~Soares-Cunha, B.~Coimbra, A.~J.
  Rodrigues, M.~Copelli, F.~S. Matias, N.~A. de~Vasconcelos, P.~V. Carelli,
  State-dependent complexity of the local field potential in the primary visual
  cortex, Physical Review E 110~(1) (2024) 014402.

\bibitem{Montani2015neuronas}
F.~Montani, R.~Baravalle, L.~Montangie, O.~A. Rosso, Causal information
  quantification of prominent dynamical features of biological neurons,
  Philosophical Transactions of the Royal Society A: Mathematical, Physical and
  Engineering Sciences 373~(2056) (2015) 20150109.

\bibitem{Montani2014}
F.~Montani, E.~B. Deleglise, O.~A. Rosso, Efficiency characterization of a
  large neuronal network: A causal information approach, Physica A: Statistical
  Mechanics and its Applications 401 (2014) 58--70.

\bibitem{deLuise2021network}
R.~De~Luise, R.~Baravalle, O.~A. Rosso, F.~Montani, Network configurations of
  pain: an efficiency characterization of information transmission, The
  European Physical Journal B 94 (2021) 1--13.

\bibitem{carlos2020anticipated}
F.-L.~P. Carlos, M.-M. Ubirakitan, M.~C.~A. Rodrigues, M.~Aguilar-Domingo,
  E.~Herrera-Guti{\'e}rrez, J.~G{\'o}mez-Amor, M.~Copelli, P.~V. Carelli, F.~S.
  Matias, Anticipated synchronization in human eeg data: Unidirectional
  causality with negative phase lag, Physical Review E 102~(3) (2020) 032216.

\bibitem{Grosse02}
I.~Grosse, P.~Bernaola-Galv{\'a}n, P.~Carpena, R.~Rom{\'a}n-Rold{\'a}n,
  J.~Oliver, H.~E. Stanley, Analysis of symbolic sequences using the
  jensen-shannon divergence, Physical Review E 65~(4) (2002) 041905.

\bibitem{Martin06}
M.~Martin, A.~Plastino, O.~Rosso, Generalized statistical complexity measures:
  Geometrical and analytical properties, Physica A: Statistical Mechanics and
  its Applications 369~(2) (2006) 439--462.

\bibitem{bavassi2019retrieval}
L.~Bavassi, C.~Forcato, R.~S. Fern{\'a}ndez, G.~De~Pino, M.~E. Pedreira, M.~F.
  Villarreal, Retrieval of retrained and reconsolidated memories are associated
  with a distinct neural network, Scientific Reports 9~(1) (2019) 784.

\bibitem{Kosem16}
A.~K{\"o}sem, A.~Basirat, L.~Azizi, V.~Van~Wassenhove, High-frequency neural
  activity predicts word parsing in ambiguous speech streams, Journal of
  neurophysiology 116~(6) (2016) 2497--2512.

\bibitem{michalareas2016alpha}
G.~Michalareas, J.~Vezoli, S.~Van~Pelt, J.-M. Schoffelen, H.~Kennedy, P.~Fries,
  Alpha-beta and gamma rhythms subserve feedback and feedforward influences
  among human visual cortical areas, Neuron 89~(2) (2016) 384--397.

\bibitem{pinheiro2023direct}
P.~Pinheiro-Chagas, F.~Chen, N.~Sabetfakhri, C.~Perry, J.~Parvizi, Direct
  intracranial recordings in the human angular gyrus during arithmetic
  processing, Brain Structure and Function 228~(1) (2023) 305--319.

\bibitem{pinheiro2024spatiotemporal}
P.~Pinheiro-Chagas, C.~Sava-Segal, S.~Akkol, A.~Daitch, J.~Parvizi,
  Spatiotemporal dynamics of successive activations across the human brain
  during simple arithmetic processing, Journal of Neuroscience.

\end{thebibliography}

\end{document}